\documentclass[submission,copyright]{eptcs}
\providecommand{\event}{QPL 2024} 

\usepackage{iftex}

\usepackage{amsmath}
\usepackage{amssymb}
\usepackage{amsthm}
\usepackage{tikz-cd}
\usepackage{graphicx}
\usepackage{bbm}

\theoremstyle{definition}
\newtheorem{defin}{Definition}[section]

\theoremstyle{remark}
\newtheorem{rmk}{Remark}[section]

\theoremstyle{remark}
\newtheorem{prop}{Proposition}[section]

\theoremstyle{definition}
\newtheorem{thm}{Theorem}[section]

\theoremstyle{remark}

\theoremstyle{remark}
\newtheorem{ex}{Example}[section]

\theoremstyle{definition}

\newcommand{\thmheader}[1]{\emph{#1}}

\newcommand{\helper}[1]{\makebox[0pt]{\hss#1\hss}}
\newcommand{\diamondminus}{%
  \sbox0{$\lozenge$}%
  \usebox0\kern-.5\wd0\helper{\raisebox{.1ex}{\scalebox{.7}[1]{$-$}}}\kern.5\wd0%
}

\newcommand{\diamondplus}{%
  \sbox0{$\lozenge$}%
  \usebox0\kern-.5\wd0\helper{\raisebox{.1ex}{\scalebox{.7}[1]{$+$}}}\kern.5\wd0%
}

\ifpdf
  \usepackage{underscore}         
  \usepackage[T1]{fontenc}        
\else
  \usepackage{breakurl}           
\fi

\newcommand{\comment}[1]{}

\title{Effect Algebras as Omega Categories}

\author{Lorenzo Perticone \quad \qquad Robin Adams
\institute{Chalmers University of Technology \\ Gothenburg, Sweden}
\email{lorenzop@chalmers.se \quad \qquad robinad@chalmers.se}
}
\def\titlerunning{Effect Algebras as Omega-categories}
\def\authorrunning{L. Perticone \& R. Adams}

\begin{document}
\maketitle

\begin{abstract}
We show that an effect algebra $X$ can be regarded as a category where morphisms $f : x \to y$ are given by triples $x \leq f \leq y$. This gives an embedding $\mathbf{EA} \to \mathbf{Cat}$. Intervals $[x, y]$ prove to be effect algebras in their own right, making $\mathbf{EA}$ embed in strict 2-categories and therefore, iterating, in strict $\omega$-categories.
We give an explicit description of the resulting $\omega$-category structure, provide concrete examples for some classes of effect algebras and provide a link to existing literature on partitions of unity in effect algebras, hence relating the construction with positive operator-valued measures.
\end{abstract}

\thanks{This research was funded by SSF, the Swedish Foundation for Strategic Research, grant number FUS21-0063}

\section*{Introduction}

Effect algebras (or equivalently D-posets) have been proposed as suitable structures for quantum logic, since they axiomatize the set of effects or 'fuzzy predicates' on a quantum system (\cite{Effect-Algebras, new-trends}). Their prevalence can be motivated by quantum theoretical arguments, and arise naturally in the study of positive semidefinite operators on Hilbert spaces.

Category theory also plays a key role in the study of quantum theory: prime examples include dagger-compact categories (\cite{ABRAMSKY2009261}) and the ZX-calculus (\cite{picturing}). While more speculative, higher category theory is showing promise for applications to quantum field theory (e.g. M. Benini and A. Schenkel's work \cite{homotopical-aqft}) and its relation with cobordisms (e.g. Freed, Hopkins, Lurie and Teleman's work \cite{tqft-compact-lie-groups}).

In this paper we present a novel link between quantum logic and (strict) higher category theory: every effect algebra gives rise to a strict $\omega$-category. We give an explicit description of such a construction, discuss how the construction applies in some well-known examples, and sketch a way to tie our work with existing literature on partitions of unity and POVMs.

\section{Background}

We start by briefly reviewing the notions of effect algebra and D-poset, their morphisms, the equivalence between the notions, and some key facts about them. As far as we know, the notion of ``generalized-abstract'' morphism of effect algebras (and likewise for D-posets) is novel.

\begin{defin}\label{eas}\thmheader{Categories of Effect Algebras (see \cite{Effect-Algebras, algebraic-quantum-logic})}

An effect algebra $(X, 1, \oplus, (-)^\dag)$ is given by:
\begin{enumerate}
\item a set $X$ together with a chosen element $1 \in X$,
\item a commutative and associative\footnote{Whenever it makes sense to require this; e.g. commutativity here means that whenever either side of $x \oplus y = y \oplus x$ is defined so is the other and they are equal.} partial binary operation $\oplus : X \times X \to X$,
\item A (total) function $(-)^\dag : X \to X$ (write $0 := 1^\dag$).
\end{enumerate}
such that 
\begin{enumerate}
\item $x \oplus x^\dag$ is always defined and equals $1$,
\item whenever $x \oplus 1$ is defined, $x = 0$,
\item whenever $x \oplus y$ is defined and equals $1$, then $y = x^\dag$.
\end{enumerate}
Given effect algebras $X$ and $Y$, we say that a function $f : X \to Y$ is:
\begin{enumerate}
\item a \emph{generalized-abstract} morphism of effect algebras if  $f(x \oplus y) = (f(x)^\dag \oplus f(0))^\dag \oplus f(y)$ whenever $x \oplus y$ is defined,
\item a \emph{generalized} morphism of effect algebras if $f(x \oplus y) = f(x) \oplus f(y)$ whenever $x \oplus y$ is defined,
\item a \emph{morphism of effect algebras} if it is a generalized morphism and $f(1) = 1$.
\end{enumerate}
Morphisms are generalized morphisms, which are in turn generalized-abstract. All three notions compose, and identities are trivially morphisms: we hence get wide-subcategory embeddings
\[\mathbf{EA} \hookrightarrow \mathbf{gEA} \hookrightarrow \mathbf{gaEA}\]
\end{defin}
In any given effect algebra $(X, 1, \oplus, (-)^\dag)$ the following hold:
\begin{enumerate}
\item $(-)^\dag$ is an involution, meaning $(x^\dag)^\dag = x$,
\item $x \oplus 0$ is always defined, and equals $x$,
\item if $x \oplus y$ is defined and equals $0$ then $x = y = 0$,
\item if $x \oplus y = x \oplus z$ then $y = z$.
\end{enumerate}
A notion equivalent to that of effect algebras is the following:
\begin{defin}\label{dpos}\thmheader{Categories of D-Posets (see \cite{D-posets, gen-dpos-ortho})}

A \emph{D-poset} $(X, \le, 1, \ominus)$ is given by:
\begin{enumerate}
\item a poset $(X, \le)$ with a top element $1$,
\item a partial binary operation $\ominus : X \times X \to X$ (write $0 := 1 \ominus 1$).
\end{enumerate}
such that, writing $X_{\le x}$ for $\{ y \in X : y \leq x\}$:
\begin{enumerate}
\item for all $x \in X$, we have $x \ominus - : X_{\le x} \to X_{\le x}$ is a (total) order-reversing involution,
\item whenever $z \le y \le x$ we have $(x \ominus z) \ominus (x \ominus y) = y \ominus z$.
\end{enumerate}
Given D-posets $X$ and $Y$, we say that a monotonic function $f : X \to Y$ is:
\begin{enumerate}
\item \emph{generalized-abstract D-monotonic} if $y \le x$ implies $f(x \ominus y) \ominus f(0) = f(x) \ominus f(y)$,
\item \emph{generalized D-monotonic} if $y \le x$ implies $f(x \ominus y) = f(x) \ominus f(y)$,
\item \emph{D-monotonic} if it is generalized D-monotonic and $f(1) = 1$.
\end{enumerate}
D-monotonic functions are generalized D-monotonic, which are in turn generalized-abstract D-monotonic. All three notions compose and identities are D-monotonic: we get wide-subcategory embeddings
\[\mathbf{DPos} \hookrightarrow \mathbf{gDPos} \hookrightarrow \mathbf{gaDPos}\]
\end{defin}
In any given D-poset $(X, \le, \top, \ominus)$ the following hold:
\begin{enumerate}
\item for all $x, y \in X$ we have $x \ominus x = y \ominus y$ which is hence $\bot$,
\item if $x \ominus y = x$ then $y = \bot$, and if $x \ominus y = \bot$ then $y = x$;
\item if $z \leq y \leq x$, we get $y \ominus z \leq x \ominus z$ and $(x \ominus z) \ominus (y \ominus z) = x \ominus y$;
\item if $y \leq x$ and $z \leq x$, then $x \ominus y = z$ iff $x \ominus z = y$;
\item if $y \leq x$ and $z \leq x \ominus y$, then $y \leq x \ominus z$ and $(x \ominus y) \ominus z = (x \ominus z) \ominus y$;
\item for all $x \in X$ the antitonic function $x \ominus -$ is an order-isomorphism;
\item for all $x \in X$ we can see $- \ominus x$ as an order-preserving (total) function.
\begin{equation*}
(- \ominus x) : X_{x \leq} \to X_{\leq (\top \ominus x)}
\end{equation*}
\end{enumerate}

\begin{thm}\label{EA-DPos-equivalence}\thmheader{Equivalence between D-Posets and Effect Algebras (see \cite{dpos-eas})}

Given an effect algebra $(X, 1, \oplus, (-)^\dag)$ we can define a D-poset $(X, \le, 1, \ominus)$ by setting $x \le y$ if there is a (necessarily unique) $z$ such that $x \oplus z = y$; in that case we define $y \ominus x := z$.

Conversely, given a D-poset $(X, \le, 1, \ominus)$ we can define an effect algebra $(X, 1, \oplus, (-)^\dag)$ by setting $x^\dag := 1 \ominus x$, and stipulating that $x \oplus y$ is defined if there is a (necessarily unique) $z$ such that $z \ominus y = x$; in that case, $x \oplus y := z$. These constructions are inverses of one another, and carry over to (generalized, abstract) morphisms, so we have isomorphisms of categories $\mathbf{EA} \cong \mathbf{DPos}$, $\mathbf{gEA} \cong \mathbf{gDPos}$ and $\mathbf{gaEA} \cong \mathbf{gaDPos}$.

More explicitly, we can define $y \ominus x := (y^\dag \oplus x)^\dag$ and $y \oplus x := 1 \ominus ((1 \ominus y) \oplus x)$.
\end{thm}
From this and the results we listed above, the following can be proven (see \cite{gen-dpos-ortho, new-trends-quantum} for 1 - 3):
\begin{enumerate}
\item[1.] $x \oplus y$ is defined iff $x \leq y^\dag$,
\item[2.] given $x \leq y \leq z$, we have $(z \ominus y) \oplus (y \ominus x) = z \ominus x$,
\item[3.] given $y \leq z$ and $x \leq z \ominus y$ we have $(z \ominus y) \ominus x = z \ominus (y \oplus x)$,
\item[4.] given $x \leq w$ and $z \leq y \leq (w \ominus x)^\dag$, we have $(w \ominus x) \oplus (y \ominus z) = ((w \ominus x) \oplus y) \ominus z$,
\item[5.] For any $x \in X$, $\oplus$ defines a morphism of effect algebras $[0, x] \times [0, x^\dag] \to X$ (cft. with \ref{intervals} and \ref{products} below). In particular, whenever well-defined, the following equality holds:
\[(w \ominus x) \oplus (y \ominus z) = (w \oplus y) \ominus (x \oplus z)\]
\end{enumerate}

Before moving forward, it's useful to mention two construction that will recur in the following. The first one is that of a special class of subobjects:

\begin{defin}\label{intervals}\thmheader{Intervals}

Fix an effect algebra $X$ and two elements $x \le y$. The interval $[x, y] := \{ z \in X | x \le z \le y \}$ can be canonically endowed with the structure of an effect algebra, that is a subobject $[x, y]$ of $X$ in the category $\mathbf{gaEA}$. In the special case $x = 0$, this is in addition a subobject in $\mathbf{gEA}$.
\begin{proof}
We give the D-poset structure. The relation is given by the restriction of $\le$; the top element is $y$ and the subtraction is given by $(a, b) \mapsto (b \ominus a) \oplus x$. Checking that subset inclusion is a generalized-abstract morphism is trivial, as is checking that it is only a generalized morphism if $a = 0$.
\end{proof}
\end{defin}

We now turn to Cartesian products in $\mathbf{gaEA}$:

\begin{defin}\label{products}\thmheader{Cartesian product of effect algebras}

Given effect algebras $X$ and $Y$, then $X \times Y$ is a D-poset under the pointwise operations
\begin{align*}
    (x_1,y_1) \leq (x_2,y_2) & \Leftrightarrow x_1 \leq x_2 \wedge y_1 \leq y_2 \\
    (x_1,y_1) \ominus (x_2,y_2) & = (x_1 \ominus x_2, y_1 \ominus y_2)
\end{align*}
with top element $(1, 1)$. Then $X \times Y$ is the Cartesian product of $X$ and $Y$ in the categories $\mathbf{EA}$, $\mathbf{gEA}$ and $\mathbf{gaEA}$.
\end{defin}

\section{Effect algebras as (higher) categories and enrichment}

We shall describe in this section a way to make an effect algebra into a category (inspired by the construction that identifies a monoid with a category with one object). These categories will be canonically enriched over a category of effect algebras: it will hence be possible to iterate the construction, yielding (strict) $(n, n)$-categories for every $n \in \mathbb{N}$. The construction will then easily generalize to $n = \infty$. 

\paragraph{Note} In category theory, it is customary to identify posets with thin categories. Since all effect algebras are D-posets (\ref{EA-DPos-equivalence}), every effect algebra is canonically a category in this way. Our construction produces a \emph{different} category from an effect algebra, and the categories produced are \emph{not} thin.

\begin{defin} \label{EAs-as-categories} \thmheader{Effect algebras as categories}

Given an effect algebra $X$, we define a category $\mathbb{B} X$ as follows:
\begin{enumerate}
\item The objects are the elements of the effect algebra: $|\mathbb{B} X| = X$.
\item Morphisms from $x$ to $y$ are given by elements between $x$ and $y$:
\[ \mathbb{B}X[x,y] = X_{xy} := \{x\} \times [x,y] \times \{y\} = \{ (x,z,y) : x \leq z \leq y \} \]
\item Identities are given by the elements themselves $\mathbf{Id}_x := (x, x, x) \in X_{x, x}$,
\item Composition is given by $\oplus$ and $\ominus$ as follows: $(y, g, z) \circ (x, f, y) := (x, (g \ominus y) \oplus f, z)$.
\end{enumerate}

It is easy to check that, given elements $x$, $y$ in an effect algebra $X$ such that $x \leq y$, we have $[x,y] \cong [0, y \ominus x]$.
(These effect algebras are isomorphic in $\mathbf{gaEA}$, $\mathbf{gEA}$ and $\mathbf{EA}$.) So we can tweak the above definition obtaining $\mathbb{B}' X$ as follows:
\begin{align*}
    |\mathbb{B}'X| & := X \\
    \mathbb{B}'X[x,y] & := \{x\} \times [\bot, y \ominus x] \times \{y\} \\
    \mathbf{Id}_x & := (x,\bot,x) \\
    (y,g,z) \circ (x,f,y) & := (x, f \oplus g, z)
\end{align*}
The categories $\mathbb{B} X$ and $\mathbb{B}' X$ are clearly isomorphic.

\end{defin}

This construction $\mathbb{B}$ is functorial: (generalized, abstract) effect algebra morphisms $f : \mathcal{X} \to \mathcal{Y}$ correspond to functors $\mathbb{B} f : \mathbb{B}\mathcal{X} \to \mathbb{B}\mathcal{Y}$.

\begin{prop} \label{D-functors} \thmheader{(Generalized, abstract) Effect algebra morphisms as functors}

The operation $\mathbb{B}$ extends to a functor $\mathbf{gaEA} \to \mathbf{Cat}$.

\begin{proof}
Given effect algebras $X$ and $Y$ and a generalized-abstract morphism $f : X \to Y$, we define a functor $\mathbb{B} f : \mathbb{B} X \to \mathbb{B} Y$
as follows
\begin{align*}
(\mathbb{B} f)(x) & := f(x) & (x \in X) \\
(\mathbb{B} f)(x, \phi, y) & := (f(x), f(\phi), f(y)) & (x \leq \phi \leq y \text{ in } X) 
\end{align*}
We check that $\mathbb{B} f$ preserves compositions. Consider a composable pair $(y, \psi, z), (x, \phi, y)$. Since its composition is $(x, (\psi \ominus y) \oplus \phi, z)$, we need to check that $f((\psi \ominus y) \oplus \phi) = (f(\psi) \ominus f(y)) \oplus f(\phi)$. We can indeed compute:
\begin{align*}
f((\psi \ominus y) \oplus \phi) &= (f(\psi \ominus y) \ominus f(0)) \oplus f(\phi) \\
&= (((f(\psi) \ominus f(y)) \oplus f(0)) \ominus f(0)) \oplus f(\phi) \\
&= (f(\psi) \ominus f(y)) \oplus f(\phi)
\end{align*}
\end{proof}
\end{prop}

\paragraph{Note}
We can similarly extend $\mathbb{B}'$ to a functor $\mathbf{gaEA} \to \mathbf{Cat}$. We give here just the action of $\mathbb{B}'f$ for a generalized abstract morphism $f$:
\[ (\mathbb{B}' f)(x, \phi, y) = (f(x), f(\phi) \ominus f(\bot), f(y)) \]
The functors $\mathbb{B}$ and $\mathbb{B}'$ are naturally isomorphic.

This functor $\mathbb{B}$ enjoys some nice properties, that we are now going to describe. First of all, it plays well with the Cartesian product we defined in \ref{products}.

\begin{rmk}\label{delooping-is-monoidal}\thmheader{$\mathbb{B}$ is monoidal}

The functor $\mathbb{B} : \mathbf{gaEA} \to \mathbf{Cat}$ is a Cartesian functor (i.e. it preserves finite products), hence a symmetric monoidal functor.

\begin{proof}
It is clear that $\mathbb{B} \mathbf{1}$ is the terminal category. The isomorphism $F : \mathbb{B}(X \times Y) \cong \mathbb{B} X \times \mathbb{B} Y$ is the identity on objects, and on morphisms is defined by
\[ F((a,b), (x,y), (c,d)) = ((a,x,c), (b,y,d))\]
where $a \leq x \leq c$ in $X$ and $b \leq y \leq d$ in $Y$.
\end{proof}
\end{rmk}

It also interacts nicely with the structure the hom-sets are canonically endowed with, which is \textit{almost} a $\mathbf{gaEA}$-enrichment. But not quite: some of the hom-sets are empty, and the emtpy set is not an effect algebra. The purpose of the next definition is to fix exactly this.

\begin{defin}\label{empty-EAs} \thmheader{Categories of possibly empty effect algebras}

We define categories $\mathbf{EA}^*$, $\mathbf{gEA}^*$, $\mathbf{gaEA}^*$ as follows:
\begin{enumerate}
\item Their objects are either the empty set or effect algebras,
\item Their morphisms are either empty maps, or (generalized, abstract) effect algebra morphisms.
\end{enumerate}
\end{defin}

Clearly, these categories inherit the Cartesian monoidal structure we introduced in \ref{products}. In what follows, we shall never mention the categories $\mathbf{EA}, \mathbf{gEA}, \mathbf{gaEA}$ again: we shall use their possibly-emtpy counterpart. We can reconstruct statements about effect algebras by composing with one (or a composition) of the inclusions in the following commutative diagram:

\begin{center}\begin{tikzcd}
\mathbf{EA} \arrow[r, hookrightarrow] \arrow[d, hookrightarrow] &
\mathbf{gEA} \arrow[r, hookrightarrow] \arrow[d, hookrightarrow] &
\mathbf{gaEA} \arrow[d, hookrightarrow] \\
\mathbf{EA}^* \arrow[r, hookrightarrow] &
\mathbf{gEA}^* \arrow[r, hookrightarrow] &
\mathbf{gaEA}^*
\end{tikzcd} \end{center}

It is trivial to notice that the functor $\mathbb{B}$ we just defined extends to a functor $\mathbb{B} : \mathbf{gaEA}^* \to \mathbf{Cat}$. We can now precisely state the second nice property of $\mathbb{B}$.

\begin{rmk}\label{delooping-is-enriched} \thmheader{$\mathbb{B}$ factrors through $\mathbf{gaEA}^*$-enriched categories}

The functor $\mathbb{B}$ sends effect algebras to $\mathbf{gaEA}^*$-enriched categories, and (generalized, abstract) effect algebra morphisms to $\mathbf{gaEA}^*$-enriched functors. In other words, it factors through the change-of-enriching-category functor stemming from the forgetful functor $\mathcal{U} : \mathbf{gaEA}^* \to \mathbf{Set}$:

\[\mathbb{B} : \mathbf{gaEA}^* \xrightarrow{\hat{\mathbb{B}}} \mathbf{Cat}_{\mathbf{gaEA}^*} \xrightarrow{\mathcal{U}_*} \mathbf{Cat}\]
\begin{proof}

Given any effect algebra $X$, we can endow the hom-sets of $\mathbb{B} X$ with the interval effect algebra structure whenever they are non-empty, since by \ref{EAs-as-categories} we clearly have $\mathbb{B}X[a, b] = \{a\} \times [a, b] \times \{b\}$, which is an effect algebra (since the interval $[a, b]$ is an effect algebra, cft \ref{intervals}) or the empty set. It is clear that identities $\mathbf{1} \to \mathbb{B}X[a, a]$ are generalized abstract effect algebra morphisms. We need to check that compositions in $\hat{\mathbb{B}} X$ is a (generalized, abstract) morphisms. This follows directly from point 5 after theorem \ref{EA-DPos-equivalence}.
\end{proof}
\end{rmk}

\begin{rmk}
The category $\mathbf{Cat}_{\mathbf{gaEA}^*}$ has finite products, and so is symmetric monoidal. Moreover, $\hat{\mathbb{B}}$ is Cartesian and hence strong symmetric monoidal.

\begin{proof}
Given $\mathcal{C}, \mathcal{D} \in \mathbf{Cat}_{\mathbf{gaEA}^*}$, the hom-sets of the product category $\mathcal{C} \times \mathcal{D}$ are the products of hom-sets and so are effect algebras by \ref{products}).

Given enriched functors $F : \mathcal{C} \to \mathcal{C}^\prime : \mathbf{Cat}_{\mathbf{gaEA}^*}$, $G : \mathcal{D} \to \mathcal{D}^\prime : \mathbf{Cat}_{\mathbf{gaEA}^*}$, it is easy to check that the functor
$ F \times G : \mathcal{C} \times \mathcal{D} \to \mathcal{C}^\prime \times \mathcal{D}^\prime $
is $\mathbf{gaEA}^*$-enriched.

The fact that $\hat{\mathbb{B}}$ is Cartesian is easy to check directly.
\end{proof}
\end{rmk}

\section{Effect Algebras as omega-categories}

We have shown that an effect algebra $X$ can be identified with a $\mathbf{gaEA}^*$-enriched category. The objects of $\mathbf{gaEA}^*$ can therefore themselves be identified with $\mathbf{gaEA}^*$-enriched categories, making $X$ into a $\mathbf{gaEA}^*$-enriched strict 2-category. This process can be iterated, identifying $X$ with a $\mathbf{gaEA}^*$-enriched strict $n$-category for all $n$, and in the limiting case, a strict $\omega$-category. In this section, we give the details of all these constructions.

Our working references for the definition of $n$- and $\omega$-categories are Street (\cite{oriented-simplexes}) and Riehl (\cite{complicial-overture})\footnote{While Street's and Riehl's definition for $n$-categories agree for $n \in \mathbb{N}$, they differ in the case $n = \omega$: Riehl's is stricter. In our case, Riehl's definition applies, but since the extra condition will be trivially verified, we shall omit any mention of it in what follows.}.

We first show how the constructions are performed from a high-level point of view, and then examine the concrete details of the $n$-categories and $\omega$-category produced.

\begin{defin}
\label{defin:B-to-the-n}
    \begin{enumerate}
\item For any monoidal category $\mathbf{V}$, we write $\mathbf{Cat}_\mathbf{V}$ for the category of $\mathbf{V}$-enriched categories.
\item Given a monoidal functor $F : \mathbf{U} \rightarrow \mathbf{V}$, we write $F_* : \mathbf{Cat}_\mathbf{U} \rightarrow \mathbf{Cat}_\mathbf{V}$ for the induced change-of-enriching-category functor.
\item Define the category $\mathbf{nCat}_{\mathbf{gaEA}^*}$ of \emph{$\mathbf{gaEA}^*$-enriched strict $n$-categories} by:
\begin{align*}
\mathbf{0Cat}_{\mathbf{gaEA}^*} & := \mathbf{gaEA}^* \\
\mathbf{(n+1)Cat}_{\mathbf{gaEA}^*} & := \mathbf{Cat}_{\mathbf{nCat}_{\mathbf{gaEA}^*}}
\end{align*}
\item For every $n$, define the functor
\[ \mathbb{B}^{(n)} : \mathbf{gaEA}^* \to \mathbf{nCat}_{\mathbf{gaEA}^*} \]
as follows:
\begin{align*}
\mathbb{B}^{(0)} & := I_{\mathbf{gaEA}^*} \\
\mathbb{B}^{(1)} & := \hat{\mathbb{B}} \\
\mathbb{B}^{(n+1)} & := \mathbf{gaEA}^* \xrightarrow{\hat{\mathbb{B}}} \mathbf{1Cat}_{\mathbf{gaEA}^*} \xrightarrow{\mathbb{B}^{(n)}_{*}} \mathbf{(n+1)Cat}_{\mathbf{gaEA}^*}
\end{align*}
\item For every $n$, define the functor $\mathbb{B}_{(n)} : \mathbf{nCat}_{\mathbf{gaEA}^*} \to \mathbf{(n+1)Cat}_{\mathbf{gaEA}^*}$ by:
\begin{align*}
    \mathbb{B}_{(0)} & := \hat{\mathbb{B}} \\
    \mathbb{B}_{(n+1)} & := (\mathbb{B}_{(n)})_*
\end{align*}
\end{enumerate}
\end{defin}

\begin{prop}
\[ \mathbb{B}^{(n)} = \mathbf{gaEA}^* \xrightarrow{\mathbb{B}_{(0)}} \mathbf{Cat}_{\mathbf{gaEA}^*} \xrightarrow{\mathbb{B}_{(2)}} \dots \xrightarrow{\mathbb{B}_{(n-1)}} \mathbf{nCat}_{\mathbf{gaEA}^*}\]

\begin{proof}
    A straightforward induction.
\end{proof}
\end{prop}

It is important to stress that in any of the steps, we are considering the \textit{category} of \textit{strict} $n$-categories\footnote{It can be shown that $\mathbf{Cat}_{\mathbf{nCat}} \simeq \mathbf{(n+1)Cat}$, making the notation consistent: see theorem 2.1.6 in \cite{complicial-overture}.}, whose $n$-fold hom-sets are all objects of $\mathbf{gaEA}^*$. With the notation just defined, consider the following diagram:

\begin{center}
\begin{tikzcd}
\mathbf{gaEA} \arrow[r, "\mathbb{B}_{(0)}"] &
\mathbf{1Cat}_{\mathbf{gaEA}^*} \arrow[r, "\mathbb{B}_{(1)}"] &
\dots \arrow[r, "\mathbb{B}_{(n-1)}"] &
\mathbf{nCat}_{\mathbf{gaEA}^*} \arrow[r, "\mathbb{B}_{(n)}"] &
\dots
\end{tikzcd}
\end{center}
We claim that the category of strict $\omega$-categories is a co-cone\footnote{Indeed, the change-of-enriching-category induced by $\mathbb{B}^\infty$ has codomain $\mathbf{Cat}_{\omega\mathbf{Cat}} \simeq \omega\mathbf{Cat}$ (cft. with 2.1.6 in \cite{complicial-overture}).}, and moreover that the component at $0$ induces all other components by iterated change-of-enriching-category constructions:
\begin{center}
\begin{tikzcd}
\mathbf{gaEA} \arrow[r, "\mathbb{B}_{(0)}"] \arrow[drr, "\mathbb{B}^\infty"'] &
\mathbf{1Cat}_{\mathbf{gaEA}^*} \arrow[r, "\mathbb{B}_{(1)}"] \arrow[dr, "\mathbb{B}^{\infty}_{*}"] &
\dots \arrow[r, "\mathbb{B}_{(n-1)}"] \arrow[d, ""] &
\mathbf{nCat}_{\mathbf{gaEA}^*} \arrow[r, "\mathbb{B}_{(n)}"] \arrow[dl, ""] &
\dots \arrow[dll, ""] \\
& & \omega\mathbf{Cat} & &
\end{tikzcd}
\end{center}

In this section we shall give a precise account of the claim in the above discussion. In order to do so we shall first describe concretely the functor $\mathbb{B}^\infty$ and the change-of-enriching-category functors it induces, then show how these make $\omega\mathbf{Cat}$ into a co-cone. In doing so, we shall explicitly describe all the finite steps $\mathbb{B}^{(n)} X$.

It will help to first explain what $k$-cells look like for small values of $k$. Given an effect algebra $X$:
\begin{itemize}
    \item the 0-cells (objects) are the elements of $X$
    \item a 1-cell from $a$ to $b$ is a triple $(a,x,b)$ with $a \leq x \leq b$
    \item a 2-cell from $(a,x,e)$ to $(a,y,e)$ is a 5-tuple $(a,x,f,y,e)$ with $a \leq x \leq f \leq y \leq e$
    \item a 3-cell from $(a,x,f,y,e)$ to $(a,x,g,y,e)$ is a 7-tuple $(a,x,f,t,g,y,e)$ with $a \leq x \leq f \leq t \leq g \leq y \leq e$
\end{itemize}
etc. The formal details are as follows:

\begin{prop}\label{EA-ncat}\thmheader{Effect Algebras as strict n-categories}

Given an effect algebra $X$, the strict $n$-category $\mathbb{B}^{(n)} X$ is as follows:
\begin{itemize}
    \item a \emph{$k$-cell} ($0 \le k \le n$) is a $2k+1$-tuple of elements of $X$, $(a_0, a_1, \ldots, a_{2k})$,
    such that $a_0 \le a_1 \le \cdots \le a_{2k}$.
    \item Given a $k+1$-cell $c = (a_0, a_1, \ldots, a_{2k+2})$,
    define its \emph{source} $s(c)$ and \emph{target} $s(c)$ (both $k$-cells) by: $s(c)$ is the result of deleting $a_k$ and $a_{k+1}$; $t(c)$ is the result of deleting $a_{k-1}$ and $a_k$. That is,
    \begin{align*}
        s(c) & = (a_0, a_1, \ldots, a_{k-1}, a_{k+2}, a_{k+3}, \ldots, a_{2k+2}) \\
        t(c) & = (a_0, a_1, \ldots, a_{k-2}, a_{k+1}, a_{k+2}, \ldots, a_{2k+2}) \enspace .
    \end{align*}
    \item Given a $k$-cell $c = (a_0, a_1, \ldots, a_{2k})$, the \emph{identity} $k+1$-cell $\mathrm{id}_c$ is defined to be the result of repeating the middle entry three times, that is:
    \[ \mathrm{id}_c = (a_0, a_1, \ldots, a_{k-1}, a_k, a_k, a_k, a_{k+1}, a_{k+2}, \ldots, a_{2k})\]
    \item \emph{Composition} Let $f$ and $g$ be $k+1$-cells with $s(g) = t(f)$. Then $f$ and $g$ must have the form
    \begin{align*}
    f & = (a_0, a_1, \ldots, a_{k-2},\ x, y, a_{k+1},\  a_{k+2}, \ldots, a_{2k+2}) \\
    g & = (a_0, a_1, \ldots, a_{k-2},\  a_{k+1}, z, w,\  a_{k+2}, a_{k+3}, \ldots, a_{2k+2})
    \end{align*}
    We define their composite to be the $k+1$-cell
    \[ g \circ f = (a_0, a_1, \ldots, a_{k-2},\  x, z \ominus a_{k+1} \oplus y, w,\ a_{k+2}, a_{k+3}, \ldots, a_{2k+2}) \]
\end{itemize}

Given a generalized abstract effect algebra morphism $f : X \rightarrow Y$, the action of $\mathbb{B}^{(n)} f : \mathbb{B}^{(n)} X \rightarrow \mathbb{B}^{(n)} Y$ is given by
\[ \mathbb{B}^{(n)} f (a_0, a_1, \ldots, a_{2k+1}) = (f(a_0), f(a_1), \ldots, f(a_{2k+1}))\]

\begin{proof}
    These details can be checked by simply following the steps in the definition \ref{defin:B-to-the-n} above. Alternatively, one can directly check this satisfies definition 2.1.3 in \cite{complicial-overture}.
\end{proof}
\end{prop}

It is now quite straightforward to define $\mathbb{B}^\infty$:

\begin{rmk}\label{EA-omega-cat}\thmheader{Effect Algebras as strict $\omega$-categories}

We have a functor $\mathbb{B}^\infty : \mathbf{gaEA} \to \omega\mathbf{Cat}$.
\begin{proof}
The definition is almost identical to that of $\mathbb{B}^{(n)}$ in \ref{EA-ncat}, we just have to omit the upper bound on the index $k$. The proof that $\mathbb{B}^\infty X$ is a strict $\omega$-category for any effect algebra $X$ requires checking three things, which are easily verified (see definition 2.1.3 in \cite{complicial-overture}):
\begin{enumerate}
\item The definition gives a reflexive globular set,
\item Given indices $0 \le i < j$ we get a category of $i$- and $j$-cells.
\item Given indices $0 \le i < j < k$, we get a strict 2-category made of $i$-, $j$- and $k$-cells.
\end{enumerate}
\end{proof}
\end{rmk}

This allows us to define the change-of-enriching-category functors
\[(\mathbb{B}^\infty)_{(n)} : \mathbf{nCat}_{\mathbf{gaEA}^*} \to \mathbf{nCat}_{\omega\mathbf{Cat}} \simeq \omega\mathbf{Cat}\]
which will play the role of the other legs of the (alleged) co-cone diagram from the previous section. The fact that it is indeed a co-cone diagram will then follow almost immediately:

\begin{rmk}\label{co-cone}\thmheader{The diagram at the end of the previous section is indeed a co-cone}

The functor $\mathbb{B}^\infty$ factors as
\[\mathbb{B}^\infty : \mathbf{gaEA}^* \xrightarrow{\mathbb{B}^{(n)}} \mathbf{nCat}_{\mathbf{gaEA}^*} \xrightarrow{(\mathbb{B}^\infty)_{(n)}} \omega\mathbf{Cat}\]

This directly implies that all compositions starting at $\mathbf{gaEA}$ and ending at $\omega\mathbf{Cat}$ agree. Moreover, since passing to the change-of-enriching-category functor commutes with composition, the same statement holds for all compositions ending at $\omega\mathbf{Cat}$, which is exactly the claim.
\end{rmk}

\section{Examples from quantum theory and logic}

We now explore a few notable examples. We start by looking at (complext) Hilbert spaces.

\begin{ex} \label{hilbert-omega-categories} \thmheader{$\omega$-categories of subspaces}

The projections $\mathbf{Proj}(\mathbb{H})$ of a Hilbert space form an effect algebra \cite{Effect-Algebras}. This effect algebra can be described in terms of subspaces of $\mathbb{H}$ since projections correspond bijectively to their image. This allows for an explicit description of $\mathbb{B}^\infty(\mathbf{Proj}(\mathbb{H}))$:

\begin{enumerate}
\item $n$-cells are sequences of (possibly non-strict) closed subspace inclusions $V_0 \subseteq \dots \subseteq V_n \subseteq \dots \subseteq V_{2n}$
\item identities are given by replicating the subspace in position $n$
\item sources are given by erasing the subspaces in position $n$ and $n+1$ (resp. targets, $n-1$ and $n$)
\item given composable\footnote{That is, $U_i = V_i \; \forall i \notin \{n-1, n, n+1\}$ and $U_{n+1} = V_{n-1}$.} $n$-cells $i \mapsto U_i$ and $i \mapsto V_i$ the composite is defined as
\[U_0 \subseteq \dots \subseteq U_{n-1} \subseteq U_n \oplus (V_n \cap V_{n-1}^{\perp}) \subseteq V_{n+1} \subseteq \dots \subseteq V_{2n}\]
\end{enumerate}
\end{ex}

This can be swiftly generalized to effects on a Hilbert space:

\begin{ex} \label{operator-omega-category} \thmheader{$\omega$-categories of effects}

The effects\footnote{That is, operators $T$ whose spectrum lies in the unit interval $\sigma(T) \subset [0, 1]$.} on $\mathbb{H}$ form an effect algebra $\mathbf{Eff}(\mathbb{H})$ \cite{Effect-Algebras}. Addition and subtraction are defined pointwise as long as they still define effects, and $T \le S$ if $S \ominus T$ is defined. In the general case, an explicit description for $\mathbb{B}^\infty(\mathbf{Eff}(\mathbb{H}))$ can be given as follows:
\begin{enumerate}
\item $n$-cells are non-decreasing sequences $T_0 \leq \dots \leq T_n \leq \dots \leq T_{2n}$
\item identities are given by replicating the operator in position $n$
\item sources and targets are given by erasing operators (as before)
\item the composition of $i \mapsto S_i$ and $i \mapsto T_i$ is given (under the same assumptions as before) by
\[S_0 \leq \dots \leq S_{n-1} \leq S_n + T_n - T_{n-1} \leq T_{n+1} \leq \dots \leq T_{2n}\]
\end{enumerate}
\end{ex}

We can say more in case all operators involved are pair-wise commuting:

\begin{rmk} \label{commuting-operators} \thmheader{$n$-cells of commuting effects}

Fix a non-decreasing sequence of commuting operators $T_0 \leq \dots \leq T_{2n}$ on a complex Hilbert space $\mathbb{H}$, and let $v_1, \dots, v_k, \dots$ be a simultaneous eigenbasis for the operators. Furthermore, let $\lambda_{i, 1}, \dots, \lambda_{i, k}, \dots$ be the corresponding eigenvalues for $T_i$. Then we can express $T_i$ as a series (where $P_k$ is the projection operator onto $\mathbf{Span}(v_k)$):
\[T_i = \sum_{k} \lambda_{i, k} P_k\]

Assume now that we have two sequences $S = S_0 \leq \dots \leq S_{2n}$ and $T = T_0 \leq \dots \leq T_{2n}$, and that all the $S_i$ and $T_i$ commute with each other. Assume moreover that the $n$-cells they represent are composable: then we can write the n-th operator in the composition as
\[(T \circ S)_n = \sum_{k} (\lambda^{S}_{n, k} + \lambda^{T}_{n, k} - \lambda^{T}_{n-1, k}) P_k\]
This means that for commuting operators, the categorical structure in play is inherited by the one stemming from the effect algebra structure on the unit interval $\mathbb{B}^\infty([0, 1])$, which we shall describe just below.
\end{rmk}

\begin{ex} \label{unit-interval-omega-category} \thmheader{$\omega$-category from the unit interval}

The unit interval $[0, 1]$ can be canonically endowed with an effect algebra structure: addition and subtraction are the usual ones (whenever the result still lies in $[0, 1]$); orthocomplementation is given by $x \mapsto 1 - x$.

This yields an $\omega$-category $\mathbb{B}^\infty([0, 1])$ whose $n$-cells are non-decreasing sequences of $2n+1$ numbers, with the usual identities, sources, targets, and composition: given composable $n$-cells $x, y$, we have

\[(x \circ y)_n = x_n + y_n - y_{n-1}\]

\end{ex}

Another important class of examples is given by Boolean algebras (in fact, every effect algebra is canonically a colimit of effect algebras stemming from Boolean algebras).

\begin{ex} \label{boolean-omega-category} \thmheader{$\omega$-category of a Boolean algebra}

Given a Boolean algebra $\mathcal{B}$, we can construct an effect algebra by defining $x \oplus y := x \vee y$ whenever $x \land y = \bot$, and $x^\dag := \neg x$; this implies that $x \ominus y = x \land \neg y$. The $\omega$-category stemming from this has the expected $n$-cells, compositions and identities; given composable $n$-cells $\{x_k\}_{0 \le k \le 2n}$ and $\{y_k\}_{0 \le k \le 2n}$, then, their composition has at position $n$ the element
\[x_n \vee (y_n \land \neg y_{n-1}) = (y_n \to y_{n-1}) \to x_n\]
\end{ex}

This has two important consequences:
\begin{enumerate}
\item Since $\sigma$-algebras are Boolean algebras, we get a functor $\mathbf{Meas}^{op} \to \omega\mathbf{Cat}$.
\item Since morphisms into $\mathbbm{1}+\mathbbm{1}$ in a topos form a Boolean algebra\footnote{Which is that of subobjects if $\mathbb{E}$ is a Boolean topos.}, we get a functor $\mathbb{E}^{op} \to \omega\mathbf{Cat}$.
\end{enumerate}

Jacobs \cite{new-trends} has defined a large class of categories called \emph{effectuses} in which the \emph{predicates} $X \to \mathbbm{2}$ on any object $X$ enjoy a canonical effect module structure (proposition 4.6 in \cite{new-trends}) and are hence effect algebras: this gives a functor $\mathbf{E}^{op} \to \omega\mathbf{Cat}$ for any effectus $\mathbf{E}$.

\comment{
\draft{\textit{TODO}} examples include, among others:
\begin{enumerate}
\item[\checkmark] Projection operators on a Hilbert space, or equivalently its closed subspaces.
\item More generally, complemented subspaces of a Banach space.
\item Even more generally, (good?) submodules of $R$-modules for $R$ a (good enough?) commutative ring.
\item[\checkmark] Bounded operators on a Hilbert space (what about Banach? or $R$-modules?).
\item[\checkmark] Boolean algebras, hence predicates in a Topos.
\item[\checkmark] Predicates in an Effectus.
\end{enumerate}
}

\section{Partitions of unity and (discrete) POVMs}

It is known in the literature that a positive operator-valued measure (POVM) on a Hilbert space $\mathbb{H}$ can be described as a morphism of effect algebras $\Sigma_A \to \mathbf{Eff}(\mathbb{H})$ that preserves countable joins, for any given measurable space $(A, \Sigma_A)$. Since finite $\sigma$-algebras are exactly the Boolean algebras, in the finite case a (discrete) POVM is just an effect algebra morphism $\mathbf{B} \to \mathbf{Eff}(\mathbb{H})$ for a finite Boolean algebra $\mathbf{B}$.

A partition of unity in an effect algebra $(X, 1, \oplus, (-)^\dag)$ is a finite multiset of elements that sum up to $1$ (for a more precise definition, we refer the reader to Lobski's work \cite{quantum-quirks}). Just as for partitions of sets, we define the \textit{refinement partial order} $\preceq$ on the set of partitions of unity in an effect algebra.

Any discrete POVM gives rise to a partition of unity: we can take the image of the atoms of the (finite, hence atomic) Boolean algebra, since by definition they sum to 1. The converse is also true: we just take the sub-multisets of a given partition of unity, which naturally make up a Boolean algebra\footnote{Some care is needed about the definition of 'sub-multiset' here: the multiset $\{a,a\}$ should have 4 sub-multisets, not 3. This can be achieved by defining a multiset over a set $X$ as an (isomorphism class of) object(s) the slice category $\mathbf{Set}/X$, and defining sub-multisets as subobjects therein.}. Moreover, the construction is 1-to-1: in what follows, we shall hence only mention partitions of unity.

The $\omega$-category $\mathbb{B} X$ induced by an effect algebra relates directly with its partitions of unity:

\begin{rmk}\label{omega-partitions} \thmheader{Partitions of unity from $\mathbb{B}^\infty$}

For every $n$-cell $\phi$ in $\mathbb{B}^\infty(\mathcal{X})$ there's a corresponding partition of unity $\mathcal{P}(\phi)$ in $\mathcal{X}$.

\begin{proof}

Given an $n$-cell $x_0 \le \dots \le x_{2n}$, we
\begin{enumerate}
\item define $y_0, \dots, y_{2n+1}$ by $y_0 = x_0$, $y_k = x_k \ominus x_{k-1}$ for $0 < k \le 2n$, and $y_{2k+1} = 1 \ominus x_{2k}$,
\item remove any $y_k$ that is equal to $0$,
\item forget the order, obtaining a multiset.
\end{enumerate}

It's trivial to check that $\mathcal{P}(\phi)$ is a partition of unity: it is summable with sum $1$ by construction.
\end{proof}
\end{rmk}

The construction just sketched is not injective: e.g. any $n$-cell gives the same partition of unity as its identity $n+1$-cell. Sources and targets relate with the refinement partial order:

\begin{rmk}\label{source-target-partition}\thmheader{Cells refine their source and target}

Given an $n$-cell $\phi$ in $\mathbb{B}^\infty(\mathcal{X})$, we have $\mathcal{P}(s(\phi)) \preceq \mathcal{P}(\phi)$ and $\mathcal{P}(t(\phi)) \preceq \mathcal{P}(\phi)$.

\begin{proof}
\[s(x_0 \le \dots \le x_{n-1} \le x_n \le x_{n+1} \le x_{n+2} \le \dots \le x_{2n}) = (x_0 \le \dots \le x_{n-1} \le x_{n+2} \le \dots \le x_{2n})\]
It is hence enough to notice that
\[x_{n+2} \ominus x_{n-1} = (x_{n+2} \ominus x_{n+1}) \oplus (x_{n+1} \ominus x_n) \oplus (x_n \ominus x_{n-1})\]
Similar considerations apply to $t(\phi)$.
\end{proof}
\end{rmk}

The relation with composition is less immediate:

\begin{rmk}\label{composition-partition}\thmheader{Compositions as refinement zig-zags}

Given composable $n$-cells $\phi, \psi$, we can ``paste'' the partitions they induce, obtaining a common refinement. This will also be a refinement of the composite $\psi \circ \phi$.

\begin{proof}
We discuss the case $n = 1$, the general case is analogous. Fix composable $1$-cells $\phi = (a \le f \le b)$ and $\psi = (b \le g \le c)$. We can first ``paste'' them, obtaining a $2$-cell $\phi \star \psi = (a \le f \le b \le g \le c)$, then build the corresponding partition
\[\mathcal{P}(\phi \star \psi) = \{a, f \ominus a, b \ominus f, g \ominus b, c \ominus g, 1 \ominus c\}\]
which refines both $\mathcal{P}(\phi)$ and $\mathcal{P}(\psi)$. It's now trivial to notice that this also refines the partition
\[\mathcal{P}(\psi \circ \phi) = \{a, (g \oplus (f \ominus b)) \ominus a, c \ominus (g \oplus (f \ominus b)), 1 \ominus c\}\]
since
\[(f \ominus a) \oplus (g \ominus b) = (g \oplus (f \ominus b)) \ominus a\]
\[(b \ominus f) \oplus (c \ominus g) = c \ominus (g \oplus (f \ominus b))\]
\end{proof}
\end{rmk}

\section{Future work}

There are broadly three different directions we see these results being expanded upon. The first one is that of increasing generality and concrete characterization:
\begin{itemize}
\item Can we say something for more general classes of linear spaces such as (complex, separable) Banach spaces and their (complemented) subspaces, or more generally ``effects'' therein?
\item Can we generalize the construction to structures such as free modules over a (nice enough) ring and their (well-behaved) free submodules?
\item Does $\mathbb{B}^\infty$ yield something known in specific cases (e.g. ordered abelian groups)
\item Does a similar construction work for Heyting algebras? If so, is there an appropriate generalization for the notion of effect algebras?
\end{itemize}
The second direction concerns exploring the (higher) categorical properties of $\mathbb{B}^n$, $\mathbb{B}^{(n)}$ and $\mathbb{B}^\infty$: 
\begin{itemize}
\item What kind of (co)limits do they preserve? Do they have an adjoint?
\item If they do, can we describe the adjoint explicitly?
\end{itemize}
The third possible direction would concern applying these constructions to established fields of research:
\begin{itemize}
\item MV-algebras can be recovered by the poset of their partitions of unity (\cite{quantum-quirks}). Does this work for a general effect algebra $X$ and $\mathbb{B}^\infty X$? What about the finite steps\footnote{One has to be careful here, because for $\mathbb{B}^{(n)}$ this is trivially true!} $\mathbb{B}^n$?
\item Linear Homotopy Type Theory (\cite{bunched-tt}) has semantics (section 3 in \cite{quantization}) in fibrations over a locally Cartesian closed $\infty$-category (e.g. in parametrized spectra \cite{parametrized-spectra}). Some such models are Boolean $\infty$-topoi, for which a variant of example \ref{boolean-omega-category} should carry through. Does this say anything interesting about such models (or about the type theory)?
\end{itemize}

\comment{
\draft{\textit{TODO}}:
\begin{enumerate}
\item Do complemented subspaces of a Banach spaces (or more generally, ``effects'' therein) make up an effect algebra?
\item What about (nice) free submodules of free modules over a (nice?) ring?
\item What does the $\omega$-category look like for initial intervals in an ordered abelian group?
\item Is the underlying (model) topos of the $(\infty, 1)$-topos of parametrized spectra Boolean? If so, what do its subobjects look like, and what about the induced $\omega$-category?
\item Can we say something about limit preservation for $\mathbb{B}$? Is it part of an adjunction between $\mathbf{gaEA}$ and $\mathbf{Cat}$? What about its iterations $\mathbb{B}^n$ and the limiting case $\mathbb{B}^\infty$?
\item Can we say more about the relation between $\mathbb{B}^\infty \mathcal{X}$ and the poset of partitions of unity in $\mathcal{X}$?
\item It's possible to reconstruct an MV-algebra $\mathcal{X}$ from its poset of partition of unity, and since $\mathbb{B}^\infty \mathcal{X}$ contains all partitions of unity and its poset structure, we can also reconstruct it from $\mathbb{B}^\infty \mathcal{X}$. Can we do it for general effect algebras by looking at the extra structure encoded in $\mathbb{B}^\infty \mathcal{X}$?
\end{enumerate}
}

\bibliographystyle{eptcs}
\bibliography{generic}
\end{document}